\documentclass[prl,twocolumn,showpacs,superscriptaddress]{revtex4-1}

\usepackage{multirow,eurosym,amssymb,amsfonts,amsmath,setspace,graphicx,color,bm,float,verbatim}

\def\be{\begin{equation}}
\def\ee{\end{equation}}

\def\bsplit{\begin{split}}
\def\nsplit{\end{split}}

\begin{document}

\title{Quantum metrology with entangled coherent states}
\date{\today}

\author{Jaewoo Joo}
\affiliation{Quantum Information Science, School of Physics and Astronomy, University of Leeds, Leeds LS2 9JT, U.K.}

\author{William J. Munro}
\affiliation{NTT Basic Research Laboratories, NTT Corporation,
3-1 Morinosato-Wakamiya, Atsugi-shi, Kanagawa 243-0198, Japan}

\affiliation{Quantum Information Science, School of Physics and
Astronomy, University of Leeds, Leeds LS2 9JT, U.K.}

\author{Timothy P. Spiller}

\affiliation{Quantum Information Science, School of Physics and
Astronomy, University of Leeds, Leeds LS2 9JT, U.K.}

\begin{abstract}
We present an improved phase estimation scheme employing
entangled coherent states and demonstrate that these states give the
smallest variance in the phase parameter in comparison to NOON,
BAT and ``optimal'' states under perfect and lossy conditions. As
these advantages emerge for very modest particle numbers, the
optical version of entangled coherent state metrology is
achievable with current technology.
\end{abstract}


\maketitle

As full quantum computing based on very large quantum resources
remains on the technological horizon for now, there is
significant current interest in quantum technologies that offer
genuine quantum advantage with much more modest quantum
resources. Quantum metrology is one field where such technologies
could emerge. Non-classical states of light can offer enhanced
imaging or spatial resolution, non-classical states of mechanical
systems could offer enhanced displacement resolution,
non-classical states of spins could enable enhanced field
resolution and entangled atoms could provide the ultimate
accuracy for clocks. Since it became known that optical quantum
states can beat the classical diffraction or shot-noise limit
\cite{Caves81}, in recent years quantum metrology has been widely
investigated in partnership with the rapid-developing field of
quantum information \cite{D08}. For example, the precision limits
of quantum phase measurements are given by the Cramer-Rao lower
limit bounded by quantum Fisher information \cite{Braunstein94}.
In the ideal quantum information version of metrology, a
maximally entangled state is viewed as the best resource for
quantum metrology, i.e. the optimal phase uncertainty of the NOON
state reaches to the Heisenberg limit and it is thus considered
for many applications (e.g. Bell's inequality tests, quantum
communication and quantum computing) \cite{Nielsen00}. However,
current quantum technologies have a long way to go to the
manipulation of many-qubit entanglement for these applications
and of course all realistic quantum technologies will be subject
to loss and decoherence. Therefore, quantum metrology utilising
very modest entangled resources and with robustness against loss
could be accessible for these applications in the much nearer
future \cite{Ralph08}, revealing a fundamental difference between
classical and quantum physics in both theory and practice.

A major research question in quantum metrology is how to implement NOON states with large particle numbers (called high NOON
states). Many successful demonstrations have shown the potential
for quantum-enhanced metrology using small NOON states
\cite{Rarity90,Afek10,Leibfried05}. However, it remains a challenge to
obtain a practical high NOON state in linear (or even non-linear)
optics. Even if high NOON states become achievable, a critical
consideration is that these states are extremely fragile to
particle loss because the resultant mixed state loses phase
information rapidly. Thus other quantum states have been studied
for improved robustness against particle loss \cite{MM08,Jess10}.
Further recent developments have shown the potential advantages of
non-linearities \cite{Caves08} and the importance of the query
complexity for quantum metrology \cite{Kok10}, concluding that
the same phase operation is required for the appropriate resource
count in different states.

In this Letter, we report that a superposition of macroscopic
coherent states { shows a noticeable improved sensitivity for} phase estimation
when compared to that for NOON states, in the region of very
modest photon or particle numbers. Taking into account the same
{\it average} particle number, an entangled coherent state (ECS)
outperforms the phase enhancement achieved by NOON states both in the lossless \cite{Peter05}, weak, moderate and high loss regimes. This advantage is also maintained over other well known quantum states used in metrology such as BAT \cite{Jess10}, and uncorrelated states. So even though
simple coherent states $|\alpha\rangle$ are known as the most
``classical-like'' quantum states \cite{Peter05},
 superpositions thereof are very useful and robust
for quantum metrology \cite{Bill02}. This phenomenon can be
understood as follows. For pure states, the ECS can be understood
as a superposition of NOON states with different photon numbers,
thus the larger photon-number NOON states make a contribution to
a better sensitivity than the average photon-number NOON state in
the ECS. For mixed states, the resultant state given by photon
loss does not depend on the number of particles lost but its loss
rate---thus this state still contains some phase information even
in the large loss rate. In order to demonstrate this perhaps
surprising phenomenon, { we suggest an implementation scheme using 
both photon-number and parity measurements} for modest ECSs which 
should be feasible with current optical technology \cite{experimentsSCS}.

We choose to compare the phase uncertainty of various quantum
states with and without loss, using the widely-accepted approach
of quantum Fisher information \cite{Braunstein94}. The
interferometric set-up generally consists of four steps. The
first is the preparation step where the input state
$|\psi^{in}_K\rangle_{12}$ is prepared in modes 1 and 2. Then, a
unitary operation $U$ in mode $2$ is applied, given by
\begin{eqnarray}
U(\phi,k) = {\rm e}^{i \phi (a^{\dag}_2 a_2)^k}
\label{eq:Phase01}
\end{eqnarray}
for phase $\phi$, order parameter of non-linearity $k$, and
creation operator $a^{\dag}_i$ in mode $i$. In this Letter we
assume $k=1$ implying that the operation $U(\phi,1)$ is a
conventional phase shifter $U(\phi)$ (although future studies
will extend this to other $k$ values). The outcome state is called
$|\psi^{out}_K\rangle_{12}= (\openone \otimes U (\phi) )
|\psi^{in}_K\rangle_{12}$. For the case of particle loss, we add
two variable beam-splitters (BSs) with loss modes $3$ and $4$
located after the phase operation. After the BSs, the mixed state
$\rho^K_{12}$ (given by tracing out the loss modes 3 and 4) is
finally measured for the estimation of phase uncertainty. A
change of transmission rate $T$ in the BSs characterises the
robustness of phase estimation for the input state against the
loss. The phase optimization given by the quantum Cram\'{e}r-Rao
bound \cite{Braunstein94} for the outcome states
$|\psi^{out}_{K}\rangle$ is described by
\begin{eqnarray}
\delta \phi_K \ge {1 \over \sqrt{\mu F_{K}^{Q}} },
\label{eq:Fisher05}
\end{eqnarray}
{ where $\mu=1$ for { a} single-shot experiment \cite{Kok10}.}
For a pure state, quantum Fisher information is given by
\begin{eqnarray}
F_{K}^{Q} = 4\big[ \langle \psi'_{K}\ |\psi'_{K}\rangle -
|\langle \psi'_{K}\ |\psi^{out}_{K}\rangle |^2 \big]
\label{eq:Fisher01}
\end{eqnarray}
for $|\psi'_{K}\rangle = \partial |\psi^{out}_{K}\rangle /\partial \phi$ \cite{WalmsleyPRL09,Jess10}.
If the outcome state is the mixed state $\rho^K_{12}$, the quantum Fisher information is given by
\begin{eqnarray}
F_{K}^{Q} = \sum_{i,j} {2\over \lambda_{i} + \lambda_{j} } | \langle \lambda_{i} | {(\partial \rho^K_{12} (\phi) / \partial \phi )} | \lambda_{j} \rangle |^2,
\label{eq:Fisher02}
\end{eqnarray}
where $\lambda_{i}$ ($| \lambda_{i} \rangle$) are the eigenvalues (eigenvectors) of $\rho^K_{12}$.

Here we focus on three important input states as
$|\psi^{in}_{K}\rangle$ ($K = N, B, C$) corresponding to NOON
$|\psi^{in}_{N}\rangle$, BAT $|\psi^{in}_{B}\rangle$
\cite{Jess10,Notation01}, and ECS \cite{Luis01} given by
\begin{eqnarray}
|\psi^{in}_{C_\alpha}\rangle_{12}
&=& {\rm e}^{- {|\alpha|^2 \over 2}} {\cal{N}}_{\alpha} \sum_{n=0}^{\infty} {\alpha^n \over n!}   \left[(a^{\dag}_{1})^n + (a^{\dag}_{2})^n\right] |0 \rangle_{1} |0\rangle_2,  \nonumber \\
&=& {\cal{N}}_{\alpha}
\big[|\alpha \rangle_{1} |0\rangle_2 + |0 \rangle_1 | \alpha \rangle_2 \big],
\label{eq:Coherent01}
\end{eqnarray}
where $|0 \rangle_i$ and $|\alpha \rangle_{i}$ are respectively Fock vacuum and
coherent states in spatial mode $i$ and
(${\cal{N}}_{\alpha} = 1 / \sqrt{2(1+{\rm e}^{-|\alpha|^2})}\,$)
\cite{Peter05}. Note that $|\psi^{in}_{C_\alpha}\rangle$ can be understood as a superposition of NOON states \cite{Luis01} (a related explanation is given in \cite{Afek10}) and the phase operation is imprinted in the outcome state
\begin{eqnarray}
|\psi^{out}_{C_\alpha}\rangle_{12}
&=& {\cal{N}}_{\alpha}
\left[|\alpha \rangle_{1} |0\rangle_2 + |0 \rangle_1 | \alpha {\rm e}^{i \phi} \rangle_2 \right]. \label{eq:Coherent02}
\end{eqnarray}
\begin{figure}[b]
\centering \hspace{-1.2cm}
\includegraphics[height=6.5cm,angle=-90]{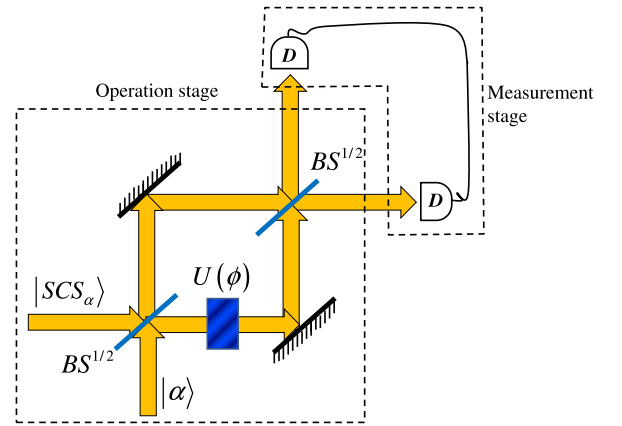}
\vspace{1cm} \caption{ Schematic illustration of an
interferometric setup for the pure ECS. Two input states
($|SCS_{\alpha}\rangle$ and $|\alpha\rangle$ are applied to the
first BS and become the ECS. After a phase shifter $U(\phi)$ in a
mode, the parity measurement is performed at the measurement
stage.} \label{fig:01}
\end{figure}
\begin{figure}[b]
\centering
\hspace{-1.5cm}
\includegraphics[height=6.5cm,angle=-90]{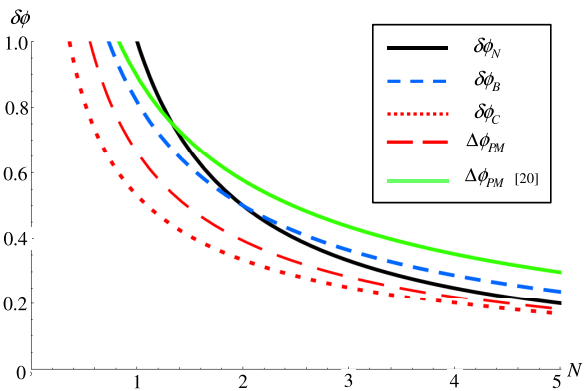}
\vspace{0.2cm}
\caption{The optimal phase estimations for NOON, BAT, and ECSs
with no particle loss are depicted in black solid and blue dashed
and red dotted lines ($ \langle n \rangle = N / 2 =
{\cal{N}}^2_{\alpha} \cdot |\alpha|^2$). Curves for NOON and BAT
states are shown as continuous for comparison, but are clearly
only defined at the appropriate integers $N$ according to
Eq.~(\ref{eq:average04}). For small $N$, $\delta \phi_N$ is
significantly bigger than $\delta \phi_C$ while $\delta \phi_N
\approx \delta \phi_C$ for large $N$. The crossover between
$\delta_{N}$ and $\delta_{B}$ at $N=2$ indicates that the NOON
and BAT states are identical. The green and red long dashed lines
show the phase estimation of the state given by Eq.~(6)
in Ref.~\cite{JPD10} and $\Delta \phi_{PM}$ above Eq.~(\ref{Parity01}), respectively.}
\label{fig:02}
\end{figure}
Considering first the situation with no loss, the optimal phase
estimation of the pure states is analytically soluble. For the
NOON and BAT states, it is equal to $\delta \phi_N \ge {1 / N}$
and $\delta \phi_B \ge {1 / \sqrt{N(N/2+1)}}$, respectively, and
for the ECS
\begin{eqnarray}
\delta \phi_C \ge {1 \over 2 \alpha {\cal{N}}_{\alpha} \sqrt{1+ \left( 1- ({\cal{N}}_{\alpha})^2 \right)  \alpha^2 }}.
\label{eq:Fisher04}
\end{eqnarray}
Taking into account equivalent resource counts for the states \cite{Bill10},
we consider the same average photon number for mode 1 given by
\begin{eqnarray}
\langle n_K \rangle = \langle  \psi^{in}_{K}|a^{\dag}_1 a_1 |\psi^{in}_{K}\rangle = {N \over 2} = {\cal{N}}^2_{\alpha} \cdot |\alpha|^2.
\label{eq:average04}
\end{eqnarray}
Then, the phase uncertainty for the ECS can be compared with
respect to $N$ for the NOON and BAT states as shown in
Fig.~\ref{fig:02}. When $N$ becomes large, $\delta \phi_{C}
\approx \delta \phi_{N} $ which indicates that the ECS becomes
approximately equivalent to the NOON state, being dominated by
the NOON amplitude at $N=|\alpha|^2$. However, interestingly,
$\delta \phi_N$ is significantly bigger than $\delta \phi_C$ for
small $N$ because $|\psi^{in}_{C}\rangle$ contains a
superposition of NOON states including $N$ values exceeding
$|\alpha|^2$. Furthermore, for small $\alpha$, the two terms in
Eq.~(\ref{eq:Coherent01}) are not orthogonal (and only tend to
being so in the large $\alpha$ limit). { The importance of non-orthogonality in a superposed single-mode state has been investigated as a quantum ruler \cite{Bill02}.} These superposition
properties enable an advantage for the coherent states at small
$|\alpha|^2$.
For a more detailed example, taking $N=4$ for the NOON and BAT
states $\langle n_{N_4} \rangle = \langle n_{B_4} \rangle = 2$ and
$\alpha = 2.0$ for the ECS (which gives a slightly lower resource
count $\langle n_{C_2} \rangle = 1.964$), the values of the
optimal phase estimation are equal to $\delta \phi_{N_4} = 0.25$,
$\delta \phi_{B_4} \approx 0.289$, and $\delta \phi_{C_2} \approx
0.205$. This indicates that even with a slight resource
disadvantage $\langle n_{C_2} \rangle < \langle n_{N_4} \rangle =
\langle n_{B_4} \rangle$, there is still a phase estimation
advantage $\delta \phi_{C_2} < \delta \phi_{N_4} < \delta
\phi_{B_4} $ (see Fig.~\ref{fig:02} around at $N=4$). This is all
very well in the zero loss regime; however, the more important
question is on the robustness of the phase { sensitivity enhancement}
in the realistic scenario of particle loss { (for instance $ \delta
\phi_{B} < \delta \phi_{N}$ for large loss \cite{Jess10} indicating BAT states are 
more sensitive in this regime).} In order to obtain
quantum Fisher information for a mixed state due to particle
loss, calculation eigenvalues and eigenvectors is required.
\begin{figure}[b]
\centering
\includegraphics[width=7.5cm]{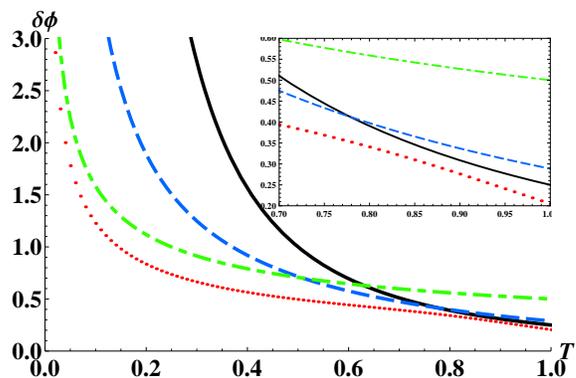}
\caption{ The graphs show the phase uncertainty with respect to
particle loss ($T$: transmisstion rate of the BSs) for four
states ($N=4$ and $\alpha=2$). { The legend of Fig.~\ref{fig:02} is 
used here while the dash-dotted green line indicates uncorrelated
states \cite{Jess10}. The ECS curve starts from $\phi_{C_2}\approx 0.205$ at $T=1$.} } \label{fig:03}
\end{figure}
{ 
From previous work \cite{Jess10}, the optimal phase
estimations for $\rho^{N}_{12}$ (NOON)  and $\rho^{B}_{12}$ (BAT) are
already known when loss is included}. { Thus, we only need to focus
on obtaining the phase estimation of the ECS
$|\psi^{out}_{C_{\alpha}}\rangle$}. { We can model such loss by two beam-splitters 
with the same transmission coefficient $T$. The total state can be written 
by $|\Psi_{C_{\alpha}} \rangle_{1234} =BS^{T}_{1,3} BS^{T}_{2,4} |\psi^{out}_{C_{\alpha}}\rangle_{12} |0\rangle_3
|0\rangle_4 $}. { Tracing out modes 3 and 4 we obtain the mixed
state $\rho^{C_{\alpha}}_{12} = \sum_{n,m=0}^{\infty} P_{n m}\, \rho_{n
m}$ where $P_{n m}={}_{1234}\langle \Psi_{C_{\alpha}}|n m \rangle_{34}
\langle n m |\Psi_{C_{\alpha}}\rangle_{1234}$ is the probability of
detecting particles $n$ in mode 3 and $m$ in mode 4 ($\rho_{n
m}$ is the resultant density operator with $P_{n m}$).
Because the case of particle loss in mode 3 (4) projects into state $|S_L\rangle_1 = |\alpha \sqrt{T} \rangle_1$ ($|S_R \rangle_2=|\alpha \sqrt{T} {\rm e}^{i \phi} \rangle_2$), the density operator can be written simply by $\rho_L = \rho_{n0} = |S_L\rangle_1 \langle S_L | \otimes |0\rangle_{2} \langle 0|$ ($\rho_R = \rho_{0m} =   |0\rangle_{1} \langle 0| \otimes |S_R\rangle_2 \langle S_R |$). Thus, the mixed state can be written in only two components given by
\begin{eqnarray}
\rho^{C_{\alpha}}_{12} = P_{00}\, \rho_{00} + P_D \, \rho_{D} ,\label{total_mixedstate02}
\end{eqnarray}
where the density operator for no particle loss is equal to
$\rho_{00} = \left| S_{00}  \right\rangle_{12}\left\langle S_{00} \right|$ ($|S_{00} \rangle = {\cal{N}}_{\alpha\sqrt{T}} (|S_L\rangle_1 |0\rangle_{2} +  |0\rangle_{1} |S_R\rangle_2 )$) and the non-normalized mixed state is $\rho_{D} = \rho_L + \rho_R $.
Note
that the resultant state is a mixture of the ECS with $\alpha\sqrt{T}$ (for no loss) and the other mixed state $\rho_{D}$ (for particle losses). The probability of no particle detection is
$P_{00} =  \left( {\rm e}^{|\alpha|^2 T} +1 \right) / \left( {\rm e}^{|\alpha|^2}+1 \right)$ and that of particle detections is $P_D = \sum_{n=1}^{\infty} \, P_{0n} = \left({\cal{N}}_{\alpha}\right)^2 \left( 1- {\rm e}^{|\alpha^2|(T-1)} \right)$.}

{ To calculate quantum Fisher information for our state $\rho^{C_{\alpha}}_{12}$, we choose $\alpha=2.0$ providing us the interesting regime for pure states and truncate the Fock basis at $n=15$ (corresponding to a maximum
error of approximately $10^{-5}$)}. { The mixed state in
Eq.~(\ref{total_mixedstate02}) is then approximately equal to
$\tilde{\rho}^{C_2}_{12}  = P_{00}\, \tilde{\rho}_{00} + \left(
\sum_{n=1}^{15} \, P_{0n} \right) \tilde{\rho}_{D}$. Using eigenvalues and eigenvectors of the
truncated density matrix $\tilde{\rho}^{C_2}_{12}$, we obtain the
optimal phase estimation of $\tilde{\rho}^{C_2}_{12}$ which we depict in
Fig.~\ref{fig:03}. The optimal phase estimation of  the entangled
coherent state clearly improves on that of NOON, BAT, and
uncorrelated states under conditions of loss, for essentially the
whole range of $T$. For $T \approx 1$, the value of the entangled
coherent state follows that of the NOON state because
$|\tilde{S}_{00}\rangle_{12}$ is the dominant factor of
$\tilde{\rho}^{C_2}_{12}$ with large probability (see the inset
in Fig.~\ref{fig:03}). However, it merges to that of the
uncorrelated state at $T\ll 1$ because $\tilde{\rho}_{D}$ makes a
major contribution in $\tilde{\rho}^{C_2}_{12}$ and is slightly
better than the uncorrelated state, ${1\over 4} (|1\rangle_1 |0\rangle_2 + {\rm e}^{i \phi} |0\rangle_1 |1\rangle_2)^{\otimes 4}$, due to phase coherence in
the projected state given by the particle detection (see the state $\rho_D$). For large $N$, $\delta \phi_C$ approaches to $\delta \phi_N$ due to $|\psi^{in}_{C}\rangle \approx |\psi^{in}_{N}\rangle$. } We further remark on comparison
with so-called ``optimal states'' \cite{WalmsleyPRL09}. Due to the concavity of
Fisher information, the engineering of optimal input states for a
known lossy rate has been considered \cite{WalmsleyPRL09}. These states effectively provide
a smooth interpolation between NOON at high $T$ and uncorrelated
at low $T$, and so ECSs also offer advantage over these states.

Having demonstrated that moderate-size ECSs offer advantage with
phase estimation, we also need to consider how such states could
be implemented in order to realise this advantage. In principle
this is achievable with current technology. There are basically
four stages: 
1) Generation of a Schr\"{o}dinger-cat state (SCS)
$|SCS_{\alpha} \rangle = {N}_{\alpha} (|\alpha\rangle + |{\rm -}
\alpha\rangle)$ for ${N}_{\alpha} = 1 / \sqrt{2(1+{\rm
e}^{-2|\alpha|^2})}$. 2) Application of $BS^{1/2}_{1,2}$ to a
coherent state $|\alpha \rangle_1$ and the Schr\"{o}dinger-cat
state $|SCS_{\alpha} \rangle_2$, with a resultant state
$|\psi^{in}_{C_{\alpha'}}\rangle = {\cal{N}}_{\alpha'} \big(|
\alpha' \rangle_1 | 0 \rangle_2 + | 0 \rangle_1 | \alpha'
\rangle_2 \big)$ where $\alpha' = \sqrt{2} \alpha$. Thus, the
state  $|SCS_{\sqrt{2}} \rangle$ is required for the input state
$|\psi^{in}_{C_2}\rangle$. According to experimental reports
\cite{experimentsSCS}, $|SCS_{\alpha} \rangle$ with $\alpha
\approx 1.5$ are already feasible in optics. 3) Application of a typical phase
shifter on mode 2, with resultant state
$|\psi^{out}_{C_{\alpha'}}\rangle$.
{ 
4) Final measurement of the variance
of the particle number in mode 2.
This measurement scheme has already been demonstrated
for mixed states with the help of the concavity of quantum Fisher
information \cite{WalmsleyPRL09}. Alternatively, a final parity measurement \cite{Gerry10} may be applicable,
such as $ \left( \Delta \phi_{PM} \right)^2 = \left( 1- \langle \Pi_{2}
\rangle^2 \right) / (\partial \langle \Pi_{2} \rangle /\partial
\phi)^2 $ given by the expectation value
\begin{eqnarray}
\langle \Pi_{2} \rangle = { 2 + {\rm e}^{-|\alpha|^2 \cos \phi}
\left({\rm e}^{-i|\alpha|^2 \sin \phi} +{\rm e}^{i|\alpha|^2 \sin
\phi} \right) \over 2+2{\rm e}^{|\alpha|^2}}~~~~
\label{Parity01}
\end{eqnarray}
for $\Pi_{2} = {\rm e}^{i\pi b^{\dag}_2 b_2}$. As shown in the red long dashed line in Fig.~\ref{fig:02}, although the
parity measurement on the pure ECS does not saturate the optimal phase estimation given by the quantum Fisher information for this state, it still beats the phase enhancement provided by the NOON state. However, for mixed states, the parity measurement is too far from the optimal POVM saturating the phase enhancement given by quantum Fisher information. }

In summary, we have evaluated analytically and numerically the
phase uncertainty of the ECS and shown that this state can outperform
the phase enhancement limit given by NOON and other states possessing the same mean particle number, for the realistic scenarios of small particle number and loss. In current optical technology, it is already feasible to obtain a travelling SCS
which is a key ingredient for the ECS. { The photon-number \cite{WalmsleyPRL09} and parity
measurement \cite{Gerry10} approaches are likely to
demonstrate an advantage over NOON and related states with current
technology}. Recent studies have investigated mixing squeezed and coherent states and
non-linearity of the phase operation \cite{Caves08,Hofmann10}. Therefore study of the effects of
squeezing variables in the SCS and
investigation of non-linear effects in the phase operation
 form very interesting future research avenues, { including imperfection studies \cite{NewJoo11}}.

We acknowledge J. J. Cooper, J. Dunningham and H. Jeong for useful discussions. We acknowledge financial support from the European Commission of the European Union under the FP7 Integrated Project
Q-ESSENCE.


\begin{thebibliography}{}
\bibitem{Notation01}
$|\psi^{in}_{N}\rangle = (|N\rangle_1 |0\rangle_2 + |0\rangle_1
|N\rangle_2)/\sqrt{2}$ and $|\psi^{in}_{B}\rangle =
\sum_{k=0}^{N/ 2} { \sqrt{(N-2k)!} \sqrt{(2k)!} \over k! ({N\over
2}- k)! \sqrt{2^{N}}}|N-2k \rangle_1  |2k \rangle_2$. The acronym BAT implies its amplitude distribution similar to bat's wings.
\end{thebibliography}
\end{document}